\documentclass[twocolumn, 10pt]{article}
\pdfoutput=1
\usepackage[utf8]{inputenc}
\usepackage{natbib}
\setcitestyle{authoryear, square, comma, aysep={,}}
\usepackage{cite}
\usepackage{amsmath,amssymb,amsfonts}
\usepackage{bm}
\usepackage{algorithmic}
\usepackage{graphicx}
\usepackage{hyperref}
\usepackage{textcomp}
\usepackage{orcidlink}
\usepackage{abstract}
\usepackage[font=footnotesize,labelfont=bf]{caption}

\usepackage{geometry}
\title{shamo: A tool for electromagnetic modeling, simulation and sensitivity analysis of the head}
\author{\small
Martin Grignard\;\orcidlink{0000-0001-5549-1861}\thanks{M. Grignard is with the GIGA CRC In-Vivo Imaging, 
University of Liège, Liège, Belgium 
(e-mail: mar.grignard@uliege.be)},
Christophe Geuzaine\;\orcidlink{0000-0001-9970-358X}\thanks{C. Geuzaine is with the Department of Electrical Engineering and Computer Science, 
University of Liège, Liège, Belgium 
(e-mail: cgeuzaine@uliege.be)},
Christophe Phillips\;\orcidlink{0000-0002-4990-425X}\thanks{C. Phillips is with the GIGA CRC In-Vivo Imaging, 
University of Liège, Liège, Belgium 
(e-mail: c.phillips@uliege.be)}
}
\date{\small Submitted on February 14, 2022}

\begin{document}

\twocolumn[
\maketitle
\begin{onecolabstract}
Accurate electromagnetic modeling of the head of a subject is of main interest in the fields of source reconstruction and brain stimulation. Those processes rely heavily on the quality of the model and, even though the geometry of the tissues can be extracted from magnetic resonance images (MRI) or computed tomography (CT), their physical properties such as the electrical conductivity are difficult to measure with non intrusive techniques. In this paper, we propose a tool to assess the uncertainty in the model parameters, the tissue conductivity, as well as compute a parametric forward models for electroencephalography (EEG) and transcranial direct current stimulation (tDCS) current distribution.\vspace{1cm}
\end{onecolabstract}
]
\saythanks

\section{Introduction}
\label{sec:introduction}

Accurate electromagnetic modeling of the head is of main interest for electrophysiological source reconstruction techniques (EEG/MEG) and brain stimulation \\(tDCS/tMS). Such modeling must capture both the spatial distribution of the tissues and their physical properties like their electrical conductivity. The former can be extracted from different anatomical imaging techniques as magnetic resonance imaging (MRI) and computed tomography (CT) but the latter is very difficult to measure \textit{in vivo} on a subject basis, even if some properties can be derived from specific MRI sequences \citep{tuch_conductivity_2001, wu_review_2018}.

Anatomically realistic models must therefore rely on values of physical parameters reported in the literature. The electrical conductivity of biological tissues have been studied since the last century \citep{burger_measurements_1943, geddes_specific_1967, gabriel_dielectric_1996a, gabriel_dielectric_1996b, gabriel_dielectric_1996c, latikka_conductivity_2001, goncalves_vivo_2003} and new methods are still published to measure them accurately for each subject \citep{akalin_acar_simultaneous_2016}. The reported values have been shown to vary both inter- and intra-subject due to temperature, health status, age, or depending on the acquisition method and environment, e.g. \textit{in vivo} vs. \textit{ex vivo} \citep{mccann_variation_2019}.

Due to the variability in the published values, the influence of the chosen parameters and geometric models on the results of the simulations have been studied for the past decades and shown to induce erroneous electric field and potential estimations \citep{haueisen_influence_1995, haueisen_influence_1997, vallaghe_global_2009, jochmann_influence_2011, montes-restrepo_influence_2014, cho_influence_2015, akalin_acar_effects_2013, wolters_influence_2006, vorwerk_influence_2019, saturnino_principled_2019}.
Errors in the localisation of the reconstructed dipoles of up to $20$ mm have been reported for basal brain locations \citep{lanfer_influences_2012, akalin_acar_effects_2013}.
Indeed, inaccuracies on the physical parameters directly result in errors in the forward models, and thus in the reconstructed sources localization or current flow. 

In order to mitigate the variability in the results, different models for the skull have been proposed \citep{sadleir_modeling_2007, dannhauer_modeling_2011, lanfer_influences_2012, montes-restrepo_influence_2014} since it acts as an electrical insulator, in EEG and tDCS, due to its low conductivity compared to the other tissues. While it is generally modeled as a single compartment, partly due to the fact that further segmenting it into spongy and compact bone is still not included in most of automated segmentation pipelines, models differentiating these two compartments have recently been proposed \citep{puonti_accurate_2020, taberna_automated_2021}. Conductivity tensors have also been considered where the radial and tangential conductivity differ \citep{fuchs_development_2007}.

The same approach applied to white matter lead to different models of its anisotropy which have been first correlated with the water self diffusion tensor derived from diffusion weighted MR images (DWI) \citep{tuch_conductivity_2001}. Later, the equilibrium, volume fraction and electrochemical models have been proposed \citep{wu_review_2018}. The influence of such anisotropy on EEG forward and inverse problems have also been studied \citep{gullmar_influence_2010, bashar_uncertainty_2010}. Conductivity tensor imaging is still an open topic with promising advances \citep{ziegler_finite-element_2014, sajib_experimental_2016, katoch_conductivity_2019}.

In the past, sensitivity analysis have been mainly conducted as the final goal of the studies. However, quantifying  uncertainty in individual models could help better understand the observed inter-subject variability in brain stimulation and source reconstruction in broader studies. This is why we introduce \textit{shamo}, a \textit{Python} open source package\footnote{\url{https://github.com/CyclotronResearchCentre/shamo}} dedicated to stochastic electromagnetic modeling of the head and sensitivity analysis of the results.

This toolbox aims at providing a unique solution for electromagnetic head modeling in both source reconstruction and brain stimulation problems. While tools already exist for each of these fields separately, for example \textit{Brainstorm} \citep{tadel_brainstorm_2011} or \textit{MNE} \citep{GramfortEtAl2013a} for EEG and \textit{SimNIBS} \citep{thielscher_field_2015} or \textit{ROAST} \citep{huang_realistic_2019} for tDCS to only name a few, \textit{shamo} offers an integrated solution. Moreover \textit{shamo} provides a single, easily extendable, API to perform mesh generation, simulation and sensitivity analysis.

To highlight the mechanisms involved in our package and demonstrate its usability and flexibility on actual cases, we apply it to the EEG forward problem and to trans-cranial direct current stimulation (tDCS) simulation. Both analyses are performed on a realistic finite element model (FEM) derived from the MIDA model \citep{iacono_mida_2015}. To evaluate the impact of different geometries for the skull, 
we build three different models, considering either one, two or thee layers for the skull, with different electrical conductivity values for the inner and outer tables for the latter.

The sensitivity is then assessed through the computation of Sobol indices \citep{sobol_global_2001}. The random input parameters considered are the values for the electrical conductivity of the tissues. To model their probability density functions, we use the truncated normal distribution published in the recent review from \citet{mccann_variation_2019} as well as a unique uniform distribution, in a worst case scenario. In the process, we compute surrogate models that, for the EEG forward problem, results in a parametric leadfield matrix that can be used to generate new forward models for any set of electrical conductivity and, for the tDCS simulation, generates a model that can compute the current density in a region of interest for the same ranges of electrical conductivity.

\section{Materials and methods}
\label{sec:methods}

In order to simulate the current flow inside the brain, a mathematical model is required. It must account for both the geometry of the tissues and their properties. This section covers the model generation, its parametrization, and sensitivity analysis.

\subsection{Finite element model generation}
\label{sec:fem}

To begin with, we focus on the geometrical aspect of the models, for which several construction methods have been proposed \citep{hallez_review_2007}:
going from a simple multi-shell sphere to a fully fledged finite element model (FEM). Two key features of FEM are its ability to capture complex shapes and to allow for anisotropic conductivity (in the form of a finite element field).
Pipelines have been developed to help researchers produce these models \citep{windhoff_electric_2013, nielsen_automatic_2018, huang_realistic_2019, vorwerk_fieldtrip-simbio_2018}.
As described by \citet{huang_realistic_2019}, most of the available solutions for automated segmentation rely either on \textit{Matlab}, through \textit{SPM}'s ``Unified Segmentation'' tool \citep{ashburner_unified_2005} and its toolboxes or on \textit{FSL} and \textit{FreeSurfer} \citep{smith_advances_2004, fischl_sequence-independent_2004}. The ensuing model generation step is generally tied to this specific segmentation method. Unfortunately this prevents geometries for atypical non-healthy subjects or simply to include other tissue types.

In \textit{shamo}, we consider a FEM approach and mesh generation but eschew the segmentation step. In effect the mesh is produced directly from a segmented volume, i.e. where voxels are labeled as being of one of any number of tissue classes. This allows us to work with more intricate structures and even to model atypical cases, e.g. with prosthesis or abnormal tissue distributions (lesions, tumor, resection,...), by using manually segmented volumes or custom automated segmentation pipelines.

For this work, we start from the multimodal imaging-based detailed anatomical model of the human head and neck (MIDA) \citep{iacono_mida_2015}: a  $350 \times 480 \times 480$ matrix of $0.5$ $\times$ $0.5$ $\times$ $0.5$ mm$^3$ voxels including $116$ different structures. Based on this model, we define three geometries with 5 to 7 different tissues (see Table \ref{tab:tissues}), differing in how the skull is modeled.

\begin{figure}[ht]
    \centering
    \includegraphics{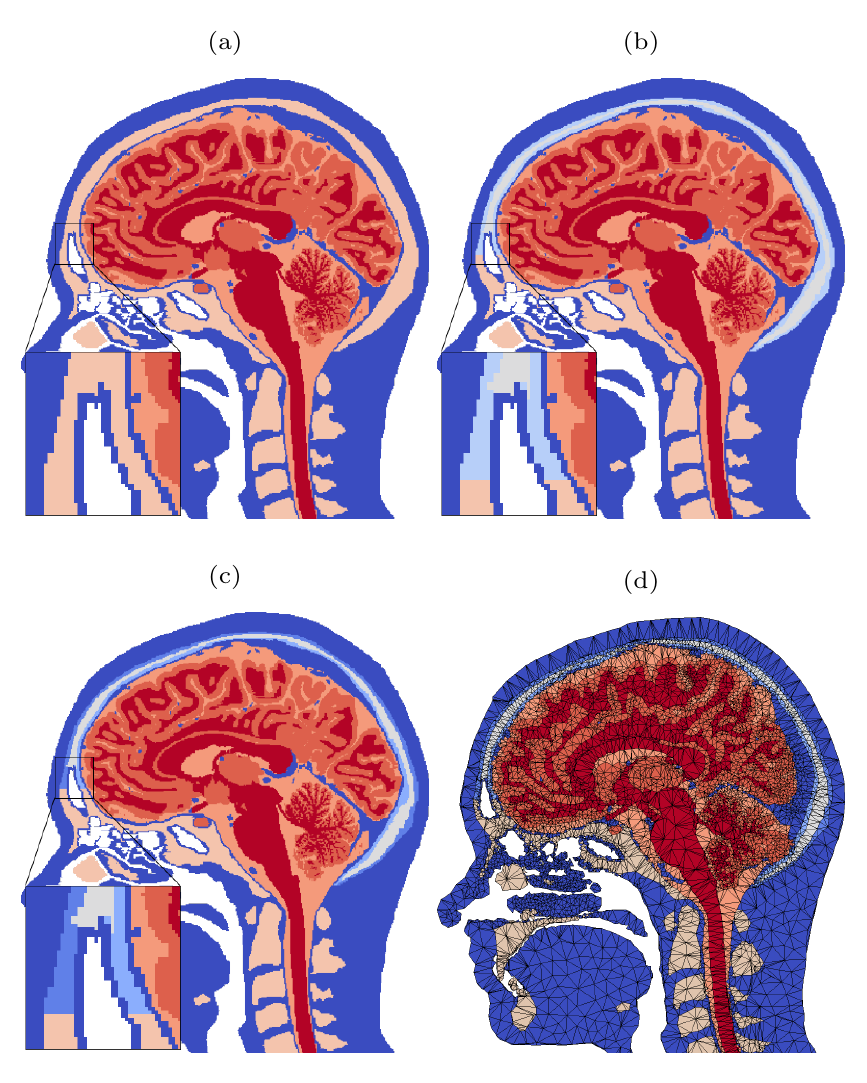}
    \caption{Sagittal cuts of (a) the segmented images for model 1 where skull is a single isotropic compartment, (b) for model 2 with spongy and compact bone differentiated, (c) for model 3 where the outer and inner tables of the skull are differentiated and (d) the mesh corresponding to model 3. In (a), (b) and (c), the lower part of the skull is the same.}
    \label{fig:models}
\end{figure}

First we merge the structures of the MIDA model to keep only the main head tissues: white matter, gray matter, cerebrospinal fluid, scalp and the different parts of the skull.
Then in model 1, the upper part of the skull is represented as a single isotropic volume; for model 2, the upper part of the skull is separated into spongia and compacta; for the model 3, the latter is further divided into outer and inner tables. Note though that the lower part of the skull is the same for all our models and is modelled as a homogeneous tissue class. The resulting models are illustrated in Figure \ref{fig:models}.

To generate the FEM tetrahedral mesh, we use \textit{CGAL} \citep{the_cgal_project_cgal_2020} with the labeled image of model 3. The resulting mesh, with $1.355 \times 10^6$ tetrahedra, then serves as a base for the other 2 models that only require the merging of some skull sub-compartments.
This merging is performed with \textit{Gmsh} \citep{geuzaine_gmsh_2009}, which is also used to annotate the mesh by specifying the names of the tissues and adding the electrodes on the scalp. For the EEG forward problem we consider the $63$ electrodes of the international 10-10 system \citep{nuwer_10-10_2018} including the fiducial markers for the nose, the left and right ears and the inion. The latter is considered as the reference for the rest of this work.
For tDCS, we use a subset of these electrodes: P3, TP9, C3, P1 and O1) where P3 is the current injector.

\begin{table*}
    \centering
    \includegraphics[scale=1]{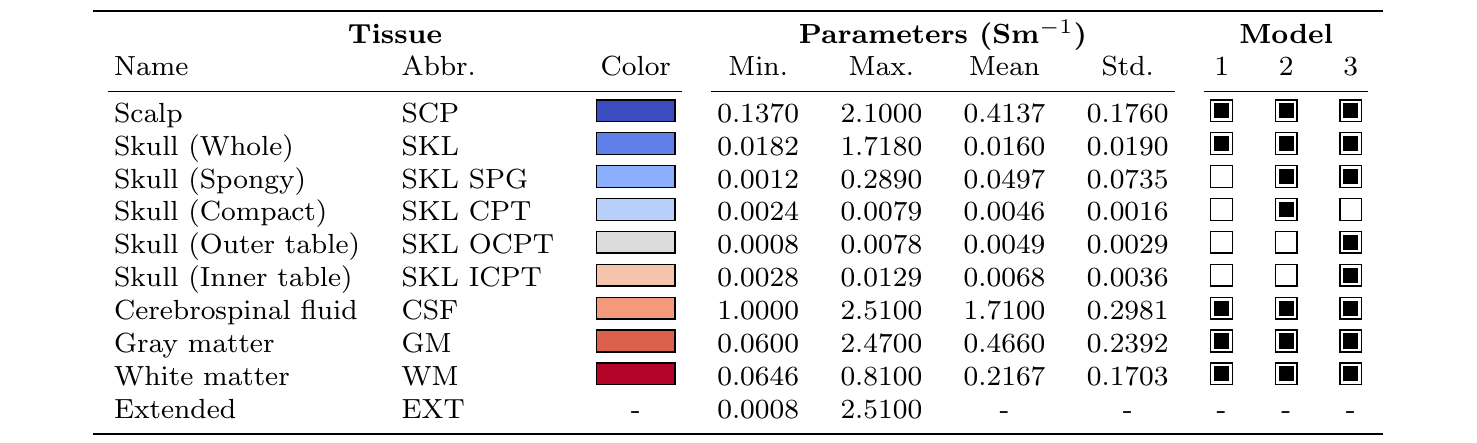}
    \caption{The tissues used in this work with the parameters of the corresponding electrical conductivity distributions from \citet{mccann_variation_2019}. The last three columns show which tissues are included in each model.}
    \label{tab:tissues}
\end{table*}

\subsection{Electromagnetic modeling}
\label{sec:electro}

Since capacitive effects can be neglected in the brain tissues for the frequency range involved in brain activity \citep{plonsey_considerations_1967}, 
the so called quasi-static approximation applies: the electromagnetic fields at time $t$ only depend on the active sources at this time.
In such conditions, Maxwell's equations reduce to a generalized Poisson equation \citep{malmivuo_bioelectromagnetism_1995, hallez_review_2007} that provides a relationship between the electric potential in any point of a volume conductor and the current sources.
We first define the current density $\bm{j}$ (Am$^{-2}$) and the source volume current density $\rho_\textrm{s}$ (Am$^{-3}$) \citep{schimpf_application_2007}. They are linked together by the expression
\begin{equation}
    \bm{\nabla} \cdot \bm{j} = \rho_\textrm{s}.
    \label{eq:poisson_1}
\end{equation}

The current density $\bm{j}$ is linearly related to the electric field $\bm{e}$ (Vm$^{-1}$) through Ohm's law
\begin{equation}
    \bm{j} = \sigma \bm{e}
    \label{eq:poisson_2}
\end{equation}
where $\sigma$, the conductivity, can be a tensor field. Anisotropy of the white matter has been shown to influence source reconstruction \citep{haueisen_influence_2002, gullmar_influence_2010} and the pipeline allows for anisotropic tissues yet, in this work, $\sigma$ is considered isotropic because no diffusion weighted images (DWI) is available in the MIDA data.
The quasi-static conditions described above allow us to write the relationship between the electric field and the electric potential field $v$ (V) as
\begin{equation}
    \bm{e} = - \bm{\nabla} v.
    \label{eq:poisson_3}
\end{equation}
Then combining Equations \eqref{eq:poisson_1}, \eqref{eq:poisson_2} and \eqref{eq:poisson_3} leads to the generalized Poisson equation:
\begin{equation}
    \bm{\nabla} \cdot (\sigma \bm{\nabla} (v)) = -\rho_\textrm{s}.
    \label{eq:poisson}
\end{equation}

Finally a homogeneous Neumann boundary condition is set on the interface between the conductor volume and the air, and a Dirichlet condition is added to set the reference electrode. We use \textit{GetDP} \citep{geuzaine_getdp_2007} as a solver for the finite element model described in section \ref{sec:fem}. \textit{GetDP} is an industrial grade solver and is at the heart of some popular tools for head electromagnetic modeling \citep{huang_realistic_2019} and our implementation of the forward problem have already been analytically validated by \citet{ziegler_finite-element_2014}.

\subsection{Electrical conductivity of tissues}
\label{sec:sigmas}

As stated in Equation \eqref{eq:poisson}, electrical conductivity plays a major role in the computation of the electric potential and the other related fields. Unfortunately, determining the exact electrical conductivity $\sigma$ (Sm$^{-1}$) of the biological tissues in a non-intrusive manner is an open issue. Multiple methods have been developed to measure it either \textit{in vitro}, \textit{ex vivo} or even \textit{in vivo} but struggle to provide an accurate and reliable value \citep{burger_measurements_1943, geddes_specific_1967, gabriel_compilation_1996, gabriel_dielectric_1996a, gabriel_dielectric_1996b, gabriel_dielectric_1996c, latikka_conductivity_2001, goncalves_vivo_2003}. \citet{mccann_variation_2019} reviewed the values acquired with different techniques and under specific conditions and derived a realistic underlying probability distribution in the form of a truncated normal distribution for the main tissues composing the head. We use these distributions with the 3 models described in Section \ref{sec:fem}, which we label Models 1a, 2a and 3a.

In addition, we define a uniform distribution spanning the whole range of the reported conductivity values for all the tissue classes, and refer to it as the extended distribution (EXT). This represents a worst case scenario with no prior information on the conductivity of the different tissues. This uniform distribution is also used with the 3 FEMs, which we label Models 1b, 2b and 3b. The range and distribution of conductivity values for the tissues considered in each model are summarized in Table \ref{tab:tissues}.

The goal of the sensitivity analysis, see Section \ref{sec:sensitivity}, is to determine the parameters that drive the variability in the results. Comparing the sensitivity to the two sets of conductivity values, realistic and extended, will allow us to assess the effect of the prior knowledge added by the truncated normal distributions, especially for the tissues with the narrowest distributions.

For the sake of clarity, we use lowercase letters to indicate a known deterministic value whereas uppercase letters refer to random values. For instance, the tissues conductivities are denoted by the vector $\bm{\sigma} = [\sigma_1, \:\sigma_2, \:\dots, \:\sigma_d]$ and $\bm{\Sigma} = [\Sigma_1, \:\Sigma_2, \:\dots, \:\Sigma_d]$ corresponds to the vector containing the random parameters modeled by the distributions.
Thus for each geometry $i$, we consider two sets of conductivity distributions: either those introduced in \citep{mccann_variation_2019}, giving $\bm{\Sigma}_\textrm{a} = [\Sigma_\textrm{WM}, \:\Sigma_\textrm{GM}, \:\dots, \:\Sigma_\textrm{SCP}]$, or the same extended distribution for each of the tissues, i.e. \\$\bm{\Sigma}_\textrm{b} = [\Sigma_\textrm{EXT}, \:\Sigma_\textrm{EXT}, \:\dots, \:\Sigma_\textrm{EXT}]$ as shown in Table \ref{tab:tissues}. 

\subsection{EEG and tDCS forward problem}
\label{sec:forward}

When carrying out EEG source reconstruction analysis, one attempts to recover the underlying brain activity inducing the observed signal at the scalp level. This electrical activity is generally modeled by one or more equivalent current dipoles characterized by their coordinates in space $\bm{r} = [r_x, \:r_y, \:r_z]$ and their dipole moment $\bm{p} = [p_x, \:p_y, \:p_z]$ (Am).  In practice, a set of discrete sources is considered rather than the full continuous volume of the gray matter. This set is called a source space and defines potential dipole locations.

The relation between the source space containing $n$ sources and the electric potential measured on $m$ electrodes at the scalp level is given by the expression
\begin{equation}
    [l] \cdot \bm{s} + \bm{\varepsilon} = \bm{v},
\end{equation}
where $\bm{s} = [p_1^{(x)}, \:p_1^{(y)}, \:p_1^{(z)}, \:\dots , \:p_n^{(x)}, \:p_n^{(y)}, \:p_n^{(z)}]^\mathsf{T}$ is the source vector and $p_j^{(k)}$ is the dipole moment of the source located in the $j$-th site along $k$-axis, \\$\bm{\varepsilon} = [\varepsilon_1, \:\dots, \:\varepsilon_m]^\mathsf{T}$ and $\bm{v} = [v_1, \:\dots, \:v_m]^\mathsf{T}$ are respectively the additive noise component vector and the vector of electrodes potentials (V), and $[l]$ is equally referred to as the "leadfield" or gain matrix. This matrix looks like this
\begin{equation}
    [l] = \begin{bmatrix}
        l_{1, 1}^{(x)} & l_{1, 1}^{(y)} & l_{1, 1}^{(z)} & \dots & l_{1, n}^{(x)} & l_{1, n}^{(y)} & l_{1, n}^{(z)} \\
        \vdots & \vdots & \vdots & \ddots & \vdots & \vdots & \vdots \\
        l_{m, 1}^{(x)} & l_{m, 1}^{(y)} & l_{m, 1}^{(z)} & \dots & l_{m, n}^{(x)} & l_{m, n}^{(y)} & l_{m, n}^{(z)}
    \end{bmatrix},
\end{equation}
where each element $l_{i,j}^{(k)}$ corresponds to the electric potential $v$ measured on the $i$-th electrode due to a current dipole with unitary dipole moment located on the $j$-th site and oriented along $k$-axis (VA$^{-1}$m$^{-1}$). This model encompasses all the geometric information and the physical properties of the head tissues. In the rest of this paper, the notation $[l(\bm{\sigma})]$ is used when highlighting the dependencies of the leadfield matrix on the values set for the electrical conductivity.

Following the method described by \citet{weinstein_lead_2000} based on the reciprocity principle, we actually have to solve the tDCS forward problem in order to generate the EEG leadfield. Indeed, to compute $[l]$ on an element basis, the reciprocity principle states that to estimate the voltage difference between two points due to a single current dipole, one needs to compute the electric field $\bm{e}$ at the coordinates of the dipole resulting from the injection of a $1$ A current $i$ between the two points, which is the definition of a tDCS forward problem.
\begin{equation}
    v_1 - v_2 = \frac{\bm{e} \cdot \bm{p}}{i}
\end{equation}
The technique then consists in setting a reference electrode, iteratively injecting a $1$ A current through the $m$ active electrodes, and estimating the electric field on the source space in the $i$-th row of the matrix $[l]$. This step of the process is achieved in \textit{GetDP} with the generalized minimal residual method (GMRES) configured with a tolerance of $10^{-8}$ and an incomplete factorization (ILU) preconditioner.

Given that the current sources should only exist in the gray matter, that the mesh for that tissue is made of about $368000$ tetrahedra, and that we have a setup of $59$ active electrodes, the whole leadfield matrix $[l]_{\textrm{full}}$ would theoretically have a size of $59 \times (3 \times 368000)$, which is too large for practical use.
Therefore we arbitrarily fix the average interval between two sources at $7.5$ mm, resulting in $2127$ sources and a leadfield matrix $[l] \in \mathbb{R}^{59 \times 6381}$ which is more manageable. This source-to-source distance influences both the the computational resources required to perform the source reconstruction, since it is directly linked to the size of the leadfield matrix, and the number of potentially reconstructed dipoles. In their paper, \citet{michel_eeg_2019} state that the definition of the spatial resolution is a sensitive problem but that increasing it does not lead to a linear increase of the accuracy. In fact, the accuracy has a limit due to the fact that the amount of information provided to the inverse problem remains constant since the number of electrodes is fixed.

\subsection{Sensitivity analysis}
\label{sec:sensitivity}

As defined by \citet{saltelli_global_2008}, sensitivity analysis is the study of how variation in the input parameters of a process influences the variation in the output. In this field, two cases are differentiated. The local sensitivity focuses on the uncertainty at a specific coordinate of the parameters space $\Omega$ whereas the global sensitivity captures the variation across the whole space.

One of the most used and studied global sensitivity analysis techniques is the computation of the so called Sobol indices \citep{sobol_global_2001, saltelli_variance_2010}. Let us consider a model $Y = Y(\bm{X})$ where $Y$ is the random output variable and $\bm{X} = [X_1, \dots, X_{n_\textrm{p}}]$ is the vector of $n_\textrm{p}$ random input variables. The first and total order Sobol indices for the $i$-th input variable, $s_i$ and $s_i^\textrm{(t)}$ are defined by
\begin{align}
    s_i & = \frac{\mathbb{V}_{X_i}(\mathbb{E}_{\bm{X}_{\setminus i}}(Y \; | \; X_i))}{\mathbb{V}(Y)}, \label{eq:sobol_first} \\
    s_i^\textrm{(t)} & = \frac{\mathbb{E}_{\bm{X} \setminus i}(\mathbb{V}_{X_i}(Y \; | \; \bm{X}_{\setminus i}) )}{\mathbb{V}(Y)}, \label{eq:sobol_total}
\end{align}
where $\bm{X}_{\setminus i}$ is the vector of all the random inputs but $X_i$, $s_i$ corresponds to the variance in the output explained by $X_i$ alone, $\mathbb{V}_{X_i}(\mathbb{E}_{\bm{X}_{\setminus i}}(Y \; | \; X_i))$ is the variance explained by the $i$-th parameter, also referred to as its main effect, and $s_i^\textrm{(t)}$ is the output variance explained by $X_i$ and all its interactions with the other input parameters.

To compute Sobol indices, we follow the method presented by \citet{saltelli_variance_2010} and implemented in the python package \textit{SALib} \citep{herman_salib_2017} that provides a way to compute both $s_i$ and $s_i^{(t)}$ from the same set of evaluations of the model, thus reducing the amount of computation required. This technique relies on $n_d$ observations $\{ (y_i^\textrm{(d)}, \bm{x}_i^\textrm{(d)}, \; i = 1, \dots, n_d \}$ where each $y_i^\textrm{(d)} = y(\bm{x}_i^\textrm{(d)})$ is the output of the model for a set of inputs $\bm{x}_i^\textrm{(d)} = [x_{i, 1}^\textrm{(d)}, \dots, x_{i, n_p}^\textrm{(d)}]$. Let us define $\bm{y}^\textrm{(d)} = [y_1^\textrm{(d)}, \dots, y_{n_d}^\textrm{(d)}]^\mathsf{T}$ the vector of outputs and $[x]^\textrm{(d)} = [\bm{x}_1^\textrm{(d)}, \dots, \bm{x}_{n_d}^\textrm{(d)}]^\mathsf{T}$ the matrix of inputs.

The matrix $[x]^\textrm{(d)}$ is built of $n_\textrm{p} + 2$ sub-matrices: $[a]$, $[b]$ and the matrices $[a_b]^{(i)}$ where all the columns are the same as in $[a]$ except the $i$-th one coming from $[b]$. All these matrices have $n_r$ rows and $n_\textrm{p}$ columns. The input vectors $\bm{x}_i^\textrm{(d)}$ composing the independent matrices $[a]$ and $[b]$ are drawn from the parameters space $\Omega$ using the Saltelli extension of Sobol quasi-random sequence \citep{sobol_distribution_1967, sobol_uniformly_1976}. Such sequences are described in section \ref{sec:surrogate}.

Based on these samples, the numerators of Equations \eqref{eq:sobol_first} and \eqref{eq:sobol_total} are computed with
\begin{equation}
    \begin{split}
        &\mathbb{V}_{X_i}(\mathbb{E}_{\bm{X}_{\setminus i}}(Y \; | \; X_i)) \\
            & \qquad = \frac{1}{n_r} \sum_{j=1}^{n_r} \bm{y}([b])_j \left( \bm{y}([a_b])_j^{(i)} - \bm{y}([a])_j \right),
    \end{split}
\end{equation}
and
\begin{equation}
    \begin{split}
        & \mathbb{E}_{\bm{X} \setminus i}( \mathbb{V}_{X_i}(Y \; | \; \bm{X}_{\setminus i}) ) \\
            & \qquad = \frac{1}{2n_r} \sum_{j=1}^{n_r} \left( \bm{y}([a])_j - \bm{y}([a_b])_j^{(i)} \right)^2.
    \end{split}
\end{equation}

\subsection{Surrogate model}
\label{sec:surrogate}

The computation of the sensitivity indices described in section \ref{sec:sensitivity} requires a large number of model evaluations. When the estimation of the actual model (here the computation of the leadfield matrix) is computationally heavy, a simpler model, referred to as the surrogate model, can be used instead. This simpler version must behave almost like if it were the real one but its evaluation should require less computing power.

Building such a model is the goal of all the supervised learning techniques. Those methods start from a set of $n_d$ observations $\bm{y}^\textrm{(d)} = [y_1^\textrm{(d)}, \:\dots, \:y_{n_d}^\textrm{(d)}]^\mathsf{T}$ of the actual model at different points of the parameters space $[x]^\textrm{(d)} = [\bm{x}_1^\textrm{(d)}, \:\dots, \:\bm{x}_{n_d}^\textrm{(d)}]^\mathsf{T}$ where $y_i^\textrm{(d)} = y(\bm{x}_i^\textrm{(d)})$ with $y(\bm{x})$ the real model. From this relatively small amount of evaluations of the model, the surrogate model $\hat{y}(\bm{x})$ is built so that $\hat{y} = \hat{y}(\bm{x}) \approx y(\bm{x})$ for any vector $\bm{x} \in \Omega$ that is not in the training set $[x]^\textrm{(d)}$.

The first step for building the surrogate model is then to draw $n_d$ vectors $\bm{x}_i^\textrm{(d)}$ to build the matrix $[x]^\textrm{(d)}$. This can be performed with various methods but here we consider quasi-random sequences. Those sequences, compared to real random sequences, take into account the previous points that have been drawn. They are used to cover the space as efficiently as possible. In section \ref{sec:sensitivity} the Saltelli extension of Sobol sequence is used to define the coordinates and here, to produce the training set for the surrogate model, we use a Halton sequence \citep{halton_efficiency_1960} as implemented in \textit{chaospy} \citep{feinberg_chaospy_2015}.

In \textit{shamo}, the generation of the surrogate model is carried out with ``Gaussian Processes Regression'' (GPR) \citep{rasmussen_gaussian_2006}. Let us define the notation for a multivariate normal distribution $\mathcal{N}(\bm{\mu}, [\gamma])$ where $\bm{\mu} = [\mu_1, \dots, \mu_{n_p}]$ is the vector of means along each axis and $[\gamma]$ is the covariance matrix where each $\gamma_{i,i}$ is the variance of the $i$-th random parameter and the elements $\gamma_{i,j}$ are the correlation between the $i$-th and the $j$-th variables.

To predict the $n_t$ values $\bm{y}^{(t)} = [y_1^{(t)}, \dots, y_{n_t}^{(t)}]^\mathsf{T}$ on the test points $[x]^{(t)}$, GPR handles the problem as Bayesian inference. Under these conditions, the learning samples are treated as random variables following a multivariate normal distribution $P(\bm{y}^\textrm{(d)} \; | \; [x]^\textrm{(d)} = \mathcal{N}(\mu^\textrm{(d)}, [\gamma]^\textrm{(d)})$. Here, the mean of this distribution is set to the mean of the learning outputs. To consider the test points, this expression becomes
\begin{equation}
    \begin{split}
        P & (\bm{y}^\textrm{(d)}, \bm{y}^{(t)} \; | \; [x]^\textrm{(d)}) \\
        & = \mathcal{N} \left( \mu^\textrm{(d)}, \begin{bmatrix}
            [\gamma]^{(t)} & [\gamma]^{(t,d)} \\
            [\gamma]^{(d, t)} & [\gamma]^\textrm{(d)} + \epsilon [i]
        \end{bmatrix} \right),
    \end{split}
\end{equation}
with $\epsilon$ an added noise.

Next, the conditional distribution $P(\bm{y}^{(t)} \; | \; \bm{y}^\textrm{(d)}, [x]^\textrm{(d)}) = \mathcal{N}(\mu^*, [\gamma]^*)$ is obtained with
\begin{align}
    \mu^* &= \mu^\textrm{(d)} + [\gamma]^{(t,d)} ([\gamma]^\textrm{(d)})^{-1} (\bm{y}^\textrm{(d)} - \mu^\textrm{(d)}),\\
    [\gamma]^* &= [\gamma]^{(t)} - [\gamma]^{(t,d)} ([\gamma]^\textrm{(d)})^{-1} [\gamma]^{(d, t)}.
\end{align}
Finally, the mean values $\mu_i^*$ and the standard deviation $\gamma_i^* = [\gamma]_{i,i}^*$ are obtained by the marginalisation of each random variable. The values $\mu_i^*$ are the predictors corresponding to the test points.

During the training step, the hyper-parameters of the kernel are optimised by maximising the log-marginal likelihood (LML) \citep{schirru_efficient_2011}. When the model outputs more than one scalar, the process can be applied separately to each of the outputs, giving one Gaussian process by output variable.
Here, we use the implementation of the GPR from \textit{scikit-learn} \citep{scikit-learn} and follow the recommendations from \citet{chen_analysis_2016}. Thus, the regression part of the GPR is set to the mean of the output variable and the kernel is obtained by the product of a constant kernel and a stationary Matérn kernel with the smoothness parameter $\nu = 2.5$, thus resulting in the covariance function
\begin{equation}
    \begin{split}
        k(\bm{x}_1, \bm{x}_2) = &\left( 1 + \frac{\sqrt{5}}{l} d(\bm{x}_1, \bm{x}_2) + \frac{5}{3l}d(\bm{x}_1, \bm{x}_2)^^2 \right) \\
        & \cdot \textrm{exp}\left( -\frac{\sqrt{5}}{l} d(\bm{x}_1, \bm{x}_2) \right).
    \end{split}
\end{equation}

\section{Applications}

We demonstrate the application of \textit{shamo} on EEG and tDCS forward problems.

\subsection{EEG forward problem}

As described in section \ref{sec:forward}, the computation of the EEG forward model is of main interest in source reconstruction but is highly dependent on the geometry and the physical properties of the tissues.

To build the surrogate model, we generate a set of leadfield matrices $[l(\bm{\sigma^{(i)})}]$ for $100$ conductivity vectors $\bm{\sigma}^{(i)}$ drawn from the parameters space $\Omega$ using a Halton sequence. This step results in a leadfield matrix where each element is actually a Gaussian process, which gives us the ability to quickly construct any new matrix $[\hat{l}(\bm{\sigma})]$ for a specific conductivity vector $\bm{\sigma}$.

The sensitivity indices defined in Equations \eqref{eq:sobol_first} and \eqref{eq:sobol_total} are only valid for a model with a single scalar output. Therefore we choose to study the sensitivity of the whole matrix to the values of $\bm{\sigma}$ with a distance measure $m(\bm{\sigma})$ relative to a reference leadfield matrix $[l]_\textrm{ref}$, obtained with a fixed $\bm{\sigma} = \bm{\sigma}_\textrm{ref}$
\begin{equation}
    m(\bm{\sigma}) = \left\lVert [l(\bm{\sigma})] - [l]_\textrm{ref} \right\rVert_\mathsf{F},
\end{equation}
where $\bm{\sigma}_\textrm{ref}$ is the mean value for each tissue (See Table \ref{tab:tissues}) and and $\left\lVert \dots \right\rVert_\mathsf{F}$ is the Frobenius norm.

\begin{figure}[ht]
    \centering
    \includegraphics{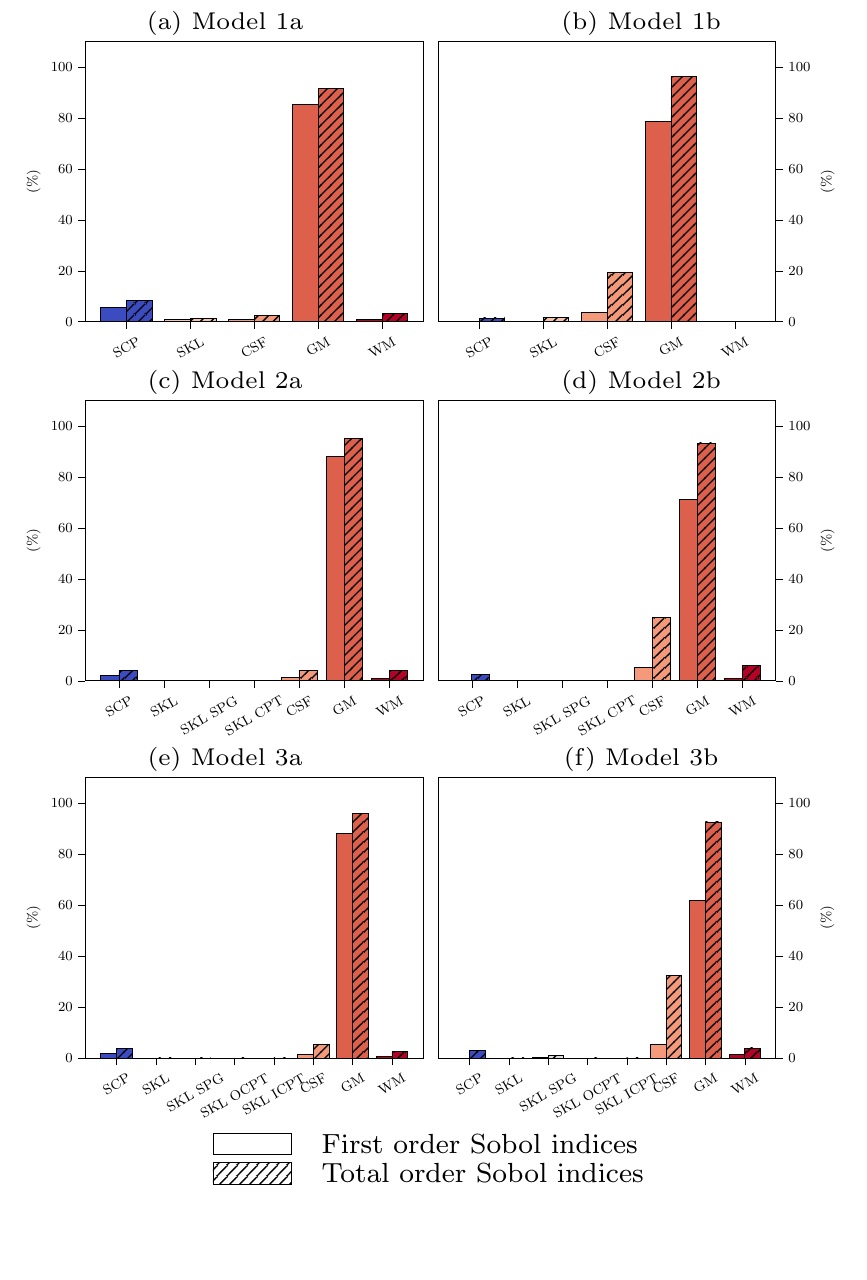}
    \caption{First (\%) and total order Sobol indices of the metric $m(\bm{\sigma})$ for each tissue of each model. In the left column, the values for $\bm{\Sigma}$ are the truncated normal distributions from \citet{mccann_variation_2019} and int the right column, the extended uniform distribution is used.}
    \label{fig:leadfield_sobol}
\end{figure}

A surrogate model $\hat{m}(\bm{\sigma})$ is thus built for this function based on the same training data as the parametric matrix introduced above. Next, the first and total order Sobol indices are computed from two sets of $40000$ evaluations of the Gaussian process for the six models of this study, as defined in Section \ref{sec:sigmas}. The resulting indices are shown in Figure \ref{fig:leadfield_sobol}.

Clearly, for both the truncated distributions and the extended uniform ones, the parameter with the largest influence on the metric is the gray matter conductivity $\sigma_{\textrm{GM}}$. Whether one uses the narrow truncated normal (models 2a/3a) or the extended uniform (models 2b/3b) distributions for the compact skull and the outer and inner tables has little influence on the Sobol indices.

Another interesting point is the increasing influence of the CSF conductivity when the uniform distribution is used. While both its first and total order Sobol indices in models 1a to 3a are very small, the value of $s_{\textrm{CSF}}^{(t)}$ for models 1b to 3b are non negligible meaning there are interactions between parameters.

\subsection{Evaluation of the electrodes potential}

\begin{figure}
    \centering
    \includegraphics{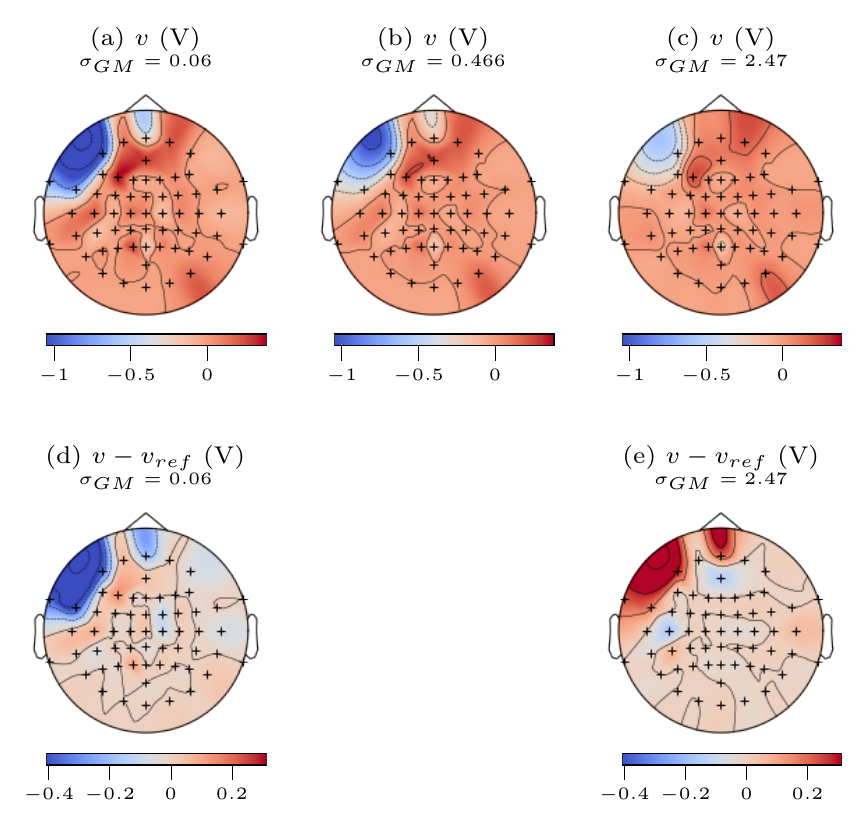}
    \caption{(a) the scalp potential (V) computed for $\sigma_\textrm{GM} = \sigma_\textrm{GM,min} =  0.06$ (Sm$^{-1}$), (b) the scalp potential (V) obtained for $\sigma_\textrm{GM} = \sigma_\textrm{GM,ref} = 0.466$ (Sm$^{-1}$) and (c) the scalp potential (V) measured with $\sigma_\textrm{GM} = \sigma_\textrm{GM,max} =  2.47$. (d) and (e) show the difference between the computed scalp potential and the reference one  (V) respectively for $\sigma_\textrm{GM} = \sigma_\textrm{GM,min}$ and $\sigma_\textrm{GM} = \sigma_\textrm{GM,max}$.}
    \label{fig:scalp_potential}
\end{figure}

To illustrate the actual effect of the sensitivity described in the above application, we calculate the electrical potential measured on the scalp due to a single left frontal source located $17.8$mm under F3. Model 3 with the $\bm{\sigma}_\textrm{ref}$ was used as a reference (Figure \ref{fig:scalp_potential}b) then the conductivity of GM was also set to the minimal and maximal value found in the literature (See Table \ref{tab:tissues}) leading to slightly modified electrical potential scalp maps (Figure \ref{fig:scalp_potential}a, c). The scalp map differences of the latter two with the reference one is shown in Figure \ref{fig:scalp_potential} d, e.

\subsection{Transcranial direct current stimulation (tDCS) simulation}

Using the same formulation as for the EEG forward problem, we can model tDCS and obtain the current density, electric potential and electric field accross the whole head.

\begin{figure}[ht]
    \centering
    \includegraphics{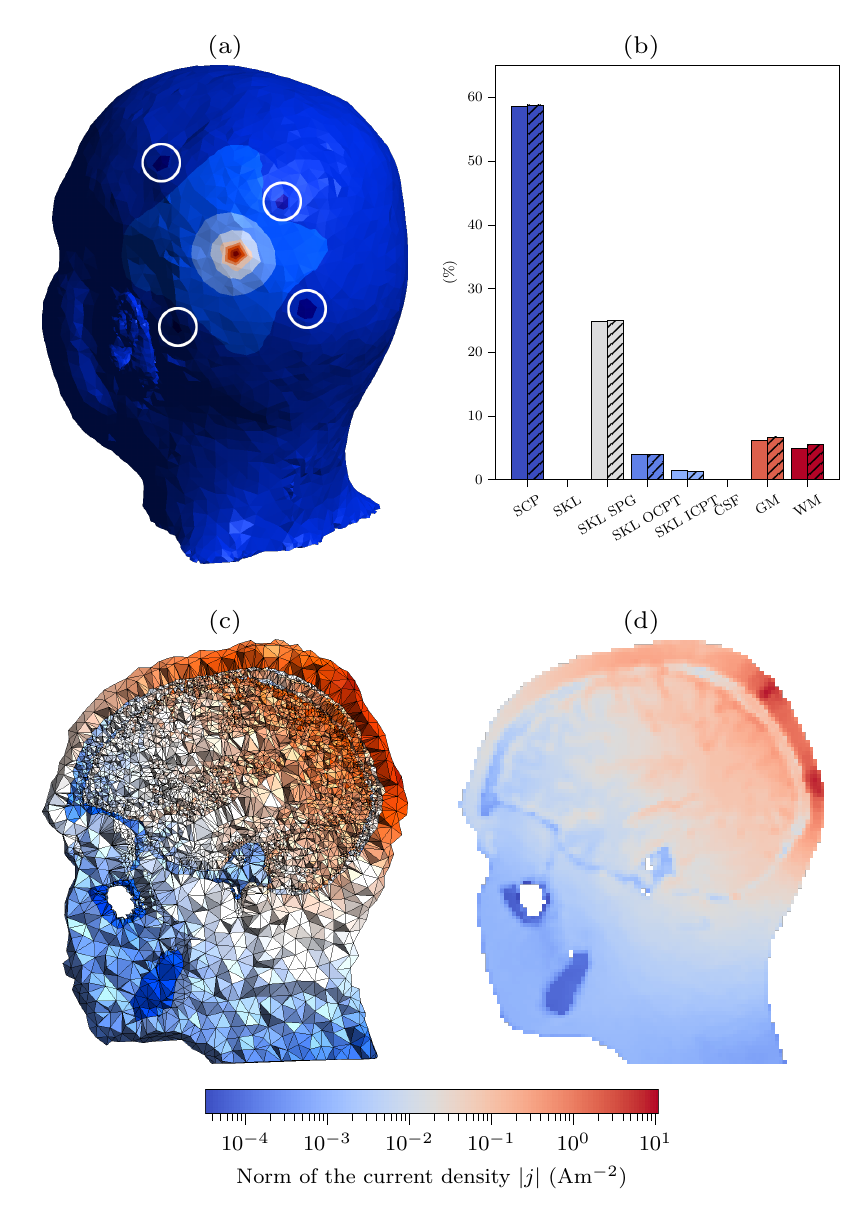}
    \caption{(a) A view of the scalp potential induced by the set electrodes surrounded in white, (b) the first (\%) and total order Sobol indices of the mean current density in the ROI for each tissue conductivity, (c) a cut of the current density inside the head resulting from the injection of current inside the reference model where $\bm{\sigma} = \bm{\sigma}_\textrm{ref}$ and (d) the same information in the form of a NIFTI file with a representation of the small mask used to compute the mean of the norm of the current density.}
    \label{fig:tdcs}
\end{figure}

To illustrate this aspect, we consider a HD-tDCS experiment where electrode P3 was set as a 4 mA injector and electrodes TP9, C3, P1 and O1 were set to ground. We used the mesh from model 3 and the truncated normal conductivity distributions as in model 3a for this simulation. As a metric to assess the sensitivity of the model with regard to the conductivity values, we chose the mean of current density norm in a small region of 368 mm$^3$ located $22.6$ mm under CP3. The results of these simulations are shown in Figure \ref{fig:tdcs}. 
As an extra feature for researchers in neuroscience, the estimated fields can also be directly exported as a standard multidimensional NifTI image.

As visible in Figure \ref{fig:tdcs}, current density is highest in the scalp tissue between the electrodes but also spreads diffusely throughout the head volume. The tissues with the highest Sobol index are the scalp, followed by the spongy compartment of the skull. This comes as no surprise as electrical current follows the path of least resistance.

\section{Discussion}
\label{sec:discussion}

The pipeline presented in this work uses several well established methods. Here, we discuss the added value of such tool and technical choices
and compare it to established solutions such as \textit{SimNIBS} \citep{thielscher_field_2015} or \textit{BrainStorm} \citep{tadel_brainstorm_2011}.

First, the generation of a realistic subject specific head model is generally a tedious task involving the segmentation of the head volume, based on MR or CT images, then the delineation of the tissue interfaces.
This step often requires further \textit{ad hoc} cleaning to ensure the surfaces are two dimensional manifold, i.e. they are completely closed and thus have an inside and an outside. Then, those surfaces must be integrated into a single mesh and the so defined volumes filled with tetrahedral elements.

Several automated pipelines are now available is the most common toolboxes \citep{thielscher_field_2015, huang_realistic_2019, nielsen_automatic_2018} using either \textit{SPM} \citep{ashburner_unified_2005} or \textit{FSL}/\textit{FreeSurfer} \citep{smith_advances_2004, fischl_sequence-independent_2004} to segment the structural images. While this greatly simplifies model generation, since it takes care of both the segmentation and meshing steps, it also prevents the user from using non standard segmentations.

Here the FEM mesh is directly built from a 3D image made with labeled voxels. This approach not only provides more control on the refinement of the final mesh but also allows the any number of tissues, even with complex configurations. In practice, this opens the path for more specific cases such as modeling prosthetic, metallic plates or any other unusual head geometry. Moreover one can freely choose the segmentation tool to use or could even proceed manually for difficult cases.

As explained previously, the model presented in this paper does not consider the anisotropy of white matter because no tensor information is provided with the original data. However any kind of field can be included in the model and handled by the solver, \textit{GetDP}, without extra burden from the point of view of the user. Besides \textit{GetDP} already powers other similar tools like \textit{SimNIBS} \citep{thielscher_field_2015} and \textit{ROAST} \citep{huang_realistic_2019} and the forward problem implementation used in \textit{shamo} have been validated by \citep{ziegler_finite-element_2014}.

Here we only considered two conductivity dependent electrostatic problems, EEG and tDCS. Nevertheless, thanks to \textit{GetDP}'s simple formalism, \textit{shamo} could be directly extended to electromagnetic modeling processes, including MEG and TMS. Indeed this only requires the definition of the equations related to the problem, which are explicitly stored in a problem file for \textit{GetDP}. 
Nowadays other tools, like \textit{SimNIBS}, also include modules for the quantification of the effect of tissue conductivity uncertainty \citep{saturnino_principled_2019}, showing the importance of such analysis for tDCS applications.
Still, beyond electrical conductivity, the sensitivity analysis could also focus on other parameters, e.g. the injected current in HD-tDCS, through python classes available in \textit{shamo} to expose the needed parameters. 
Importantly, by definition, the Sobol indices rely on a single scalar output function, whose choice is thus critical but also lets the user focus on any feature of interest. Thereby \textit{shamo}'s flexible implementation allows one to define his own processes and sensitivity analysis.

Regarding the surrogate models, we decided to use Gaussian process because they provide information on the confidence over the solution through the standard deviation on each predicted point. This can be used to obtain more in-depth understanding of the model. Such regressor also has the advantage of not requiring huge amounts of training data. Considering the fact that an evaluation of the actual model can take several hours, reducing the number of observations can drastically lower the computation time required. 
To further reduce this time, the tool provides an easy way to evaluate each sample point separately, thus allowing the use of high-performance computing (HPC) equipment like computer clusters. In the present work, observations were computed by batches of $100$, each on a single core on the C\'{E}CI clusters\footnote{\url{http://www.ceci-hpc.be/}}.


Overall the goal of \textit{shamo} is to provide a single tool to perform three major steps, namely FEM creation, model estimation, and sensitivity analysis. This operated with few dependencies that are all established, in an open source software working out of the box on any major operating system or on HPC platforms.

\section{Conclusion}
\label{sec:conclusion}

In this paper, we presented a python pipeline for accurate electromagnetic modeling of the head which allows for sensitivity analysis and surrogate model building, bringing together similar features as some established software, such as \textit{SimNIBS} \citep{thielscher_field_2015} or \textit{ROAST} \citep{huang_realistic_2019} for tDCS and \textit{Brainstorm} \citep{tadel_brainstorm_2011} or \textit{MNE} \citep{GramfortEtAl2013a} for EEG, unified with a single API. This tool, called \textit{shamo} \citep{grignard_shamo_2021}, and the full documentation \citep{grignard_shamo_documentation_2021} are available on Github under GPL-v3 license. A set of examples are also available in the form of jupyter notebooks.

We showed a use-case for the EEG forward problem where a parametric leadfield matrix is computed and can then be used to generate any new matrix for a specific set of tissue conductivity values and another application to tDCS where the current density in a certain region is obtained and can be studied with regard to the electrical sensitivity. Those are only two possible applications but \textit{shamo} could easily be extended to magnetic stimulation or TMS.

Considering the abstraction level of the tool and the outcome that can be obtained from it, one can use our tool to perform finite element analysis and sensitivity analysis without having to dig into those fields, letting the user employ the toolset of his choice for further analysis. \textit{shamo} could be used in various studies to assess the sensitivity of the results to some parameters or to build parametric models for complex physical fields that, otherwise, should be evaluated every time a new value is needed.

\section*{Data availability}

The data that support the findings of this study are available from the IT'IS foundation\footnote{\url{https://itis.swiss/virtual-population/regional-human-models/mida-model/}} but restrictions apply to the availability of these data, which were used under licence\footnote{\url{https://itis.swiss/assets/Downloads/VirtualPopulation/License_Agreements/LicenseAgreementMIDA.pdf}} for the current study, and so are not publicly available. Data are however available from the authors upon reasonable request and with permission of the IT'IS foundation.

\section*{Information sharing}

The source code of \textit{shamo} is available on Github\footnote{\url{https://github.com/CyclotronResearchCentre/shamo}} under GPLv3 license and is fully documented\footnote{\url{https://cyclotronresearchcentre.github.io/shamo/index.html}}. It can be installed from PyPI\footnote{\url{https://pypi.org/project/shamo/}}. In addition, \textit{jupyter} notebooks are also available on Github\footnote{\url{https://github.com/CyclotronResearchCentre/shamo-tutorials}} and show how to conduct similar studies.

\section*{Acknowledgements}

MG and CP are supported by the Fonds de la Recherche Scientifique de Belgique (F.R.S.-FNRS), the former under Grant No. EOS 30446199, Belgium.

Computational resources have been provided by the Consortium des \'{E}quipements de Calcul Intensif (C\'{E}CI), funded by the Fonds de la Recherche Scientifique de Belgique (F.R.S.-FNRS) under Grant No. 2.5020.11 and by the Walloon Region, Belgium.

%
\section*{Conflict of interest}

The authors declare that they have no conflict of interest.

{\footnotesize
\bibliography{references}
\bibliographystyle{plainnat}}

\end{document}